\documentclass[11pt,a4paper]{article}
\usepackage{graphicx}
\usepackage{amsmath,amssymb}
\usepackage{amsfonts}
\usepackage{eucal}
\usepackage{amsthm}
\usepackage[title]{appendix}
\numberwithin{equation}{section}

\def\a{\alpha}
\def\b{\beta}
\def\c{\gamma}
\def\d{\delta}
\def\e{\varepsilon}
\def\f{\varphi}
\def\g{\psi}

\def\i{\mbox{\raisebox{.5ex}{$\chi$}}}

\def\l{\lambda}
\def\m{\mu}
\def\n{\nu}

\def\s{\sigma}
\def\t{\tau}
\def\x{\xi}

\def\z{\zeta}

\newcommand{\Op}{\mathrm{Op}}

\def\re{\mathbb{R}}
\def\co{\mathbb{C}}
\def\ze{\mathbb{Z}}
\def\na{\mathbb{N}}

\def\pa{\partial}

\renewcommand{\Re}{\mathrm{Re}}
\renewcommand{\Im}{\mathrm{Im}}

\newcommand{\supp}{\mathrm{supp}}

\newcommand{\Ker}{\mathrm{Ker}}

\newcommand{\norm}[1]{\| #1 \|}
\newcommand{\bignorm}[1]{\bigl\| #1 \bigr\|}

\newcommand{\bigpare}[1]{\bigl(#1\bigr)}
\newcommand{\biggpare}[1]{\biggl(#1\biggr)}

\newcommand{\bigbrac}[1]{\bigl[#1\bigr]}

\newcommand{\bigset}[2]{\bigl\{#1\bigm|#2\bigr\}}

\newcommand{\jap}[1]{\langle #1 \rangle}

\newcommand{\bigabs}[1]{\bigl| #1 \bigr|}

\newcommand{\beq}{\begin{equation}}
\newcommand{\eeq}{\end{equation}}
\newcommand{\ba}{\begin{align}}
\newcommand{\ea}{\end{align}}

\newtheorem{thm}{Theorem}[section]
\newtheorem{lem}[thm]{Lemma}

\theoremstyle{definition}

\newtheorem{ass}{Assumption}

\theoremstyle{remark}
\newtheorem{rem}{Remark}[section]

\title{A remark on the essential self-adjointness for Klein-Gordon type operators}

\author{Shu Nakamura
\thanks{Department of Mathematics,
Faculty of Sciences,
Gakushuin University,
1-5-1, Mejiro, Toshima, Tokyo
171-8588, Japan. 
Email: shu.nakamura@gakushuin.ac.jp}
\and
Kouichi Taira
\thanks{%
Department of Mathematical Sciences, Ritsumeikan University, 1-1-1 NojiHigashi, 
Kusatsu, 525-8577 Japan. 
Email: ktaira@fc.ritsumei.ac.jp }}

\begin{document}
\maketitle

\begin{abstract}
Here we discuss a new simplified proof of the essential self-adjointness for formally  
self-adjoint differential operators of real principal type, previously proved by 
Vasy (2020) and Nakamura-Taira (2021). For simplicity, here we discuss 
the second order cases, i.e., Klein-Gordon type operators only.
\end{abstract}

\section{Introduction}

We consider the second order operator of the form 
\[
P=\sum_{j,k=1}^n D_j g^{jk}(x)D_k +\frac12 \sum_{j=1}^n (D_j u_j(x)+u_j(x)D_j)+u_0(x),
\]
on $L^2(\re^n)$, where $D_j =-i\frac{\pa}{\pa x_j}$, $j=1,\dots, n$, and $n\geq 2$. 
We suppose all the coefficients are real-valued $C^\infty$ functions. 
The top order coefficients $\{g^{jk}(x)\}$ is a Lorentzian cometric, and hence 
we suppose it is non-degenerate for all $x\in\re^n$. Moreover we suppose 
it is asymptotically flat, i.e., there is a non-degenerate matrix $\{g_0^{jk}\}$ such that 
$g^{jk}(x)\to g_0^{jk}$ as $|x|\to\infty$. More specifically, we suppose 

\begin{ass}\label{ass-a}
For all $j,k$, $g^{jk}(x)$, $u_j(x)$, $x\in\re^n$, are real-valued smooth functions. Moreover, 
there exists $0<\m<1$ such that for any $\a\in \ze_+^n$
\begin{align*}
\bigabs{\pa_x^\a(g^{jk}(x)-g_0^{jk})} &\leq C_\a \jap{x}^{-\m-|\a|},\quad x\in\re^n,\ j,k=1,\dots,n, \\
\bigabs{\pa_x^\a u_j(x)} &\leq C_\a \jap{x}^{-\m-|\a|}, \quad x\in\re^n,\ j=0, 1,\dots, n, 
\end{align*}
with some $C_\a>0$, where $\jap{x}=(1+|x|^2)^{1/2}$. 
\end{ass}

We also need the null-nontrapping condition. We write the principal symbol by 
\[
p_2(x,\x)= \sum_{j,k=1}^n g^{jk}(x)\x_j\x_k. 
\]
We denote $\exp(tH_{p_2})$ be the 
Hamilton flow on $\re^{2n}$ generated by the symbol $p_2(x,\x)$, and we write 
\[
(y(t,x_0,\x_0), \eta(t,x_0,\x_0))= \exp(tH_{p_2})(x_0,\x_0)
\]
for $t\in\re$, $(x_0,\x_0)\in\re^{2n}$. 

\begin{ass}[Null non-trapping condition] \label{ass-b}
If $(x_0,\x_0)\in p_2^{-1}(\{0\})$ and $\x_0\neq 0$, then 
$|y(t,x_0,\x_0)|\to\infty$ as $|t|\to\infty$. 
\end{ass}

Now we can state our main result. 

\begin{thm}\label{thm-main} 
Suppose Assumptions~\ref{ass-a} and \ref{ass-b}. Then $P$ is essentially 
self-adjoint on $C_0^\infty(\re^n)$. 
\end{thm}

The main purpose of this note is to simplify the argument of \cite{NT}. 
Here we use a semiclassical analytic method, and it significantly simplify the argument, especially 
concerning justifications of formal argument using the Yosida approximation. 
The local regularity of the eigenfunction (see Section \ref{sec-global-regularity}) is 
proved in this paper using the H\"ormander's propagation of singularities theorem, 
whereas in \cite{NT} it is proved using the microlocal smoothing property proved in \cite{N}, 
which also shorten the proof. 
The constructions of observables which are decreasing along classical trajectories 
are refined here to make the proof more transparent (Section~\ref{sec-uncoming-estimates}).

The self-adjointness of the Klein-Gordon type operators on spacetime does not obviously have 
physical importance. One of the motivation comes from the construction of the Feynman propagator, 
which is important in th construction of quantum field theory. The Feynman propagator may be formally 
defined as the boundary value of the resolvent $(P-z)^{-1}$ on the Minkowski spacetime. 
Duistermaat-H\"ormander \cite{DH} defines the Feynman propagator on various spacetimes as an inverse with a certain wavefront condition (see \cite[Definition 1.1]{GW1}) and showed the existence of an approximate inverse with the wave front condition. However, the existence of the Feynman propagator (that is, the actual inverse) had not been known since then.
 In \cite{GHV, GW1, GW2}, the existence of the Feynman propagator is shown on an asymptotically Minikowski spacetime under the null non-trapping condition. Independently, Derezi\'nski and Siemssen considered the existence of the Feynman propagators on more general spacetimes in \cite{DS1, DS2, DS3}. 
In particular, they attempt to define the Feynman propagator as the boundary value of the resolvent of the spacetime 
operator, and they conjectured the following: The wave operator $P$ is essentially self-adjoint on $C_0^{\infty}$ and the Feynman propagator defined in \cite{DS1,DS2} or \cite{GW1, GW2} coincides with the boundary value of the resolvent (\cite[Conjecture 8.3]{DS3}). For asymptotically Minkowski spacetimes, the first part of this problem is solved in \cite{Va} and \cite{NT} and the second part is proved in \cite{T2}. We address the first part in 
this paper.  

Recently, Dang and Wrochna (\cite{DW1, DW2}) studied spectral geometry of scattering Lorentzian spaces. In \cite{DW1}, it is shown that the scalar curvature of scattering Lorentzian spaces is related to the integral kernel of a power of the outgoing resolvent. 
In order to justify the power $(P-i\e)^{-\a}$, the essential self-adjointness of $P$ is assumed in \cite{DW1}. 

The essential self-adjointness of non-elliptic operators also attracted attention in another context.
Colin de Verdi\'ere and Bihan showed that on generic compact Lorentzian surfaces 
the wave operator is not essentially self-adjoint on the space of smooth functions (\cite{CB}). 
Moreover, they conjectured that for a symmetric differential operator on a compact manifold, 
the completeness of the Hamiltonian flow is equivalent to the essential self-adjointness. 
This conjecture is solved for real principal type operators on the one-dimensional torus in \cite{T1}.

This paper is organized as follows: In Section~\ref{sec-preliminaries}, we introduce notations, 
and prepare several basic tools. In Section~\ref{sec-uncoming-estimates}, we show that microlocal 
Sobolev-type smoothness of eigenfunctions in the incoming region. 
The global regularity of the eigenfunction is proved in Section~\ref{sec-global-regularity}, 
and then Sobolev-type smoothness of eigenfunctions in the outgoing region is proved in 
Section~\ref{sec-outgoing-estimates}. Finally we prove our main theorem in Section~\ref{sec-proof-of-main-theorem}. 
Several technical lemmas and estimates are proved in Appendices.


\section{Preliminaries}\label{sec-preliminaries}
\subsection{Symbol classes and the quantization}

We use the following symbol class: For $k,\ell\in \re$, we write 
\begin{align*}
S^{k,\ell}=S(\jap{x}^{\ell}\jap{\x}^k,g), 
\end{align*}
where $S(m,g)$ denotes the H\"ormander's symbol class (\cite[\S 18.4]{Ho})
with $g=dx^2/\jap{x}^2+d\x^2/\jap{\x}^2$. 
Namely, $a\in S^{k,\ell}$ if for any $\a,b\in\ze_+^n$ there is $C_{\a\b}>0$ such that 
\[
\bigabs{\pa_x^\a\pa_\x^\b a(x,\x)}\leq C_{\a\b}\jap{x}^{\ell-|\a|}\jap{\x}^{k-|\b|}, \quad x,\x\in\re^n.
\]
We consider semiclassical symbols, i.e., symbols with semiclassical parameter $h>0$, 
and its semiclassical Weyl quantization: 
For $a=a(h;x,\x)$, 
\begin{align*}
\Op_h(a)\f(x) &=(2\pi h)^{-n}\iint e^{i(x-y)\cdot\x/h}a\bigpare{h;\tfrac{x+y}{2},\x} \f(y)dyd\x.
\end{align*}
and we denote $\Op:=\Op_1$.
We denote the Poisson bracket by 
\[
\{a,b\}=\frac{\pa a}{\pa \x}\cdot\frac{\pa b}{\pa x}-\frac{\pa a}{\pa x}\cdot\frac{\pa b}{\pa \x}
=\frac{d}{dt}\exp(tH_a)b\Big|_{t=0}
\]
for functions $a(x,\x)$ and $b(x,\x)$ on $\re^{2n}$. 
We denote the weighted Sobolev space $H^{s,t}(\re^n)$ defined by
\[
H^{s,t}(\re^n)=\jap{x}^{-t}\jap{D_x}^{-s} \bigbrac{L^2(\re^n)}.
\]

\subsection{The first reduction}
In order to show the essential self-adjointness of a symmetric operator $P$, it is sufficient to show 
$\Ker(P^*-z_\pm)=\{0\}$ for some $z_\pm\in\co$, $\pm \Im(z_\pm)>0$. 
We concentrate on the case $z=z_+$ in the following. The other case is similar. 
Let $\g\in \Ker(P^*-z)$, then it implies
\beq\label{eq-P-z-psi}
\g\in L^2(\re^n), \quad (P-z)\g=0 \text{ in the distribution sense}.
\eeq
Our theorem is proved if \eqref{eq-P-z-psi} implies $\psi=0$. 
The first step of the proof is remark that it follows if we know $\psi$ is sufficiently 
good function. 

\begin{lem}\label{lem-the-first-reduction}
If $\psi$ satisfies \eqref{eq-P-z-psi} and $\psi\in L^2(\re^n)\cap H^{1,-1}(\re^n)$ then $\psi=0$.
\end{lem}

We give the proof in Appendix~\ref{app-0}. 

\begin{rem}
Actually, the condition is relaxed to $\psi\in L^2(\re^n)\cap H^{1/2,-1/2}(\re^n)$, and 
this condition is used in  Vasy \cite{Va} and Nakamura-Taira \cite{NT}, but this condition is 
sufficient for our purpose, and more elementary to prove. 
\end{rem}

Thus it suffices to show $\psi\in H^{1,-1}(\re^n)$ from \eqref{eq-P-z-psi}.
In fact, we shall show $\psi\in H^{N, -\c}(\re^n)$ with any $N,\c>0$. 

\subsection{Basic commutator estimate}\label{subseccommutator}
We use the following simple commutator estimate as the basic tool in the proof of 
$\psi\in H^{N, -\c}(\re^n)$. Let $B$ be an $h$-pseudodifferential operator, and 
suppose we have the following operator inequality: 
\beq\label{eq-basic-commutator-estimate}
i[B^*B, P] \geq \frac{c}{h}B^* \jap{x}^{-1} B -\tilde B^* \jap{x}^{-1} \tilde B -E^*E,
\eeq
where $c>0$ and $E$ is another pseudodifferential operator. 
Then by simple algebraic computations, we obtain 
\begin{align}
&\frac{c}{2h}\bignorm{\jap{x}^{-1/2}B\f}^2 +2(\Im z)\norm{B\f}^2 \nonumber  \\
&\qquad \leq \frac{2h}{c}\bignorm{\jap{x}^{1/2}B(P-z)\f}^2 + \norm{\jap{x}^{-1/2}\tilde B\f}^2 +\norm{E\f}^2
\label{eq-basic-commutator-inequality}
\end{align}
for $\f\in\mathcal{S}(\re^n)$ and $\norm{\cdot}=\norm{\cdot}_{L^2(\re^n)}$. 
We will apply this argument in the following steps, with various 
remainder terms $\tilde B$ and $E$ and $\Im z>0$. Moreover, when $0<\c<\frac{1}{2}$, 
$B,\tilde B$ and $E\in \bigcap_{m\in \re} \Op S^{m,\c}$, the inequality $(\ref{eq-basic-commutator-inequality})$ holds for $\f\in L^2(\re^n)$ with $(P-z)\f\in H^{0,\frac{1}{2}+\c}(\re^n)$. We also note that the eigenfunction $\g$ defined in $(\ref{eq-P-z-psi})$ satisfies these conditions.

These computations are standard, but we give the proof in Appendix~\ref{app-1} for the completeness. 


\section{Incoming estimates}\label{sec-uncoming-estimates}

\subsection{Incoming observable}\label{subsec:Inc}

\begin{figure}[t]
\centering
\includegraphics{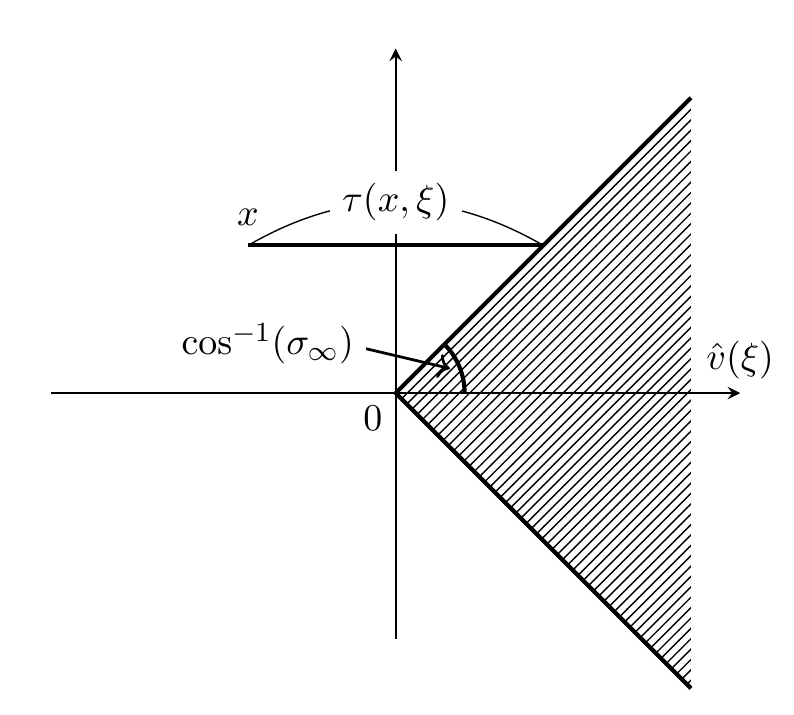}
\caption{The definition of $\t(x,\x)$}
  \label{fig:lengthtau}
\end{figure}

We use an operator of the following form: 
We write 
\[
v(\x)=\pa_\x p_0(\x), \quad \hat v(\x)=\frac{v(\x)}{|v(\x)|}, \quad \hat x =\frac{x}{|x|}, \quad 
\b(x,\x)=\hat x\cdot \hat v(\x), 
\]
and we set $0<\c\ll \m$, e.g., $\c=\m/10$, and we set $0<\s_\infty<1$, 
which is close to $1$, e.g., $\s_\infty=9/10$. 
Then we set
\beq\label{eq-def-tau}
\t(x,\x) = |x|\bigpare{c_0\sqrt{1-\b(x,\x)^2} -\b(x,\x)}, \quad c_0=\frac{\s_\infty}{\sqrt{1-\s_\infty^2}}
\eeq
for $(x,\x)$ such that $\b(x,\x)\leq \s_\infty$. The function
$\t(x,\x)$ is the length of the line segment $\bigset{x+t\hat v(\x)}{t\geq 0}$ inside 
$\bigset{(x,\x)}{\b(x,\x)\leq \s_\infty}$ (see Figure \ref{fig:lengthtau}).  
%
%

Now we set 
\[
b_-(x,\x) = \t(x,\x)^\c \z_-(x,\x), 
\]
where we specify the cut-off functions $\z_-(x,\x)$ in the following. 
We note the weight function $\t(x,\x)^\c$ is homogeneous of order $\c$ 
with respect to $x$, and order $0$ with respect to $\x$. 
We note this weight is increasing in the incoming directions, and 
the cut-off functions $\z_-(x,\x)$  localize this operator in the (microlocally) 
incoming region. The cut-off functions $\z_-(x,\x)$ also eliminate 
the singularity of this weight. 
For $0<\d\ll 1$ and $0<\s'<\s<\s_\infty$, we set
\[
\z_-(x,\x)=\z_-(\d,\s,\s',R;x,\x)=\z_1(x/R,\x)\z_2(x,\x)\z_3(x,\x), 
\]
where each function $\z_j(x,\x)$ has a different role. 

We choose a smooth functions $\i_1\in C^\infty(\re)$ such that 
\[
\i_1(s)=\begin{cases} 1, \quad & s\leq -1, \\
0, \quad & s\geq 0,\end{cases}
\]
$0\leq \i_1(s)\leq 1$, $\i_1'(s)\leq 0$ for $s\in\re$, and $0<\i_1(s)<1$ 
We also set
\[
\i_2(s)=\i_1(s-2)\i_1(2-s)
\]
so that
\[
\i_2(s)=\begin{cases} 1, \quad & |s|\leq 1, \\
0, \quad & |s|\geq 2.\end{cases}
\]

\begin{description}
\item[1. Cut-off to outside of parabola.] 
\[
\z_1(x,\x)= \i_1\bigpare{x_\x^\parallel-\tfrac12 |x^\perp_\x|^2+1}
\]
where
\[
x^\parallel_\x =x\cdot\hat v(\x), \quad x^\perp_\x =x-(x\cdot\hat v(\x))\hat v(\x)
\]
for $x,\x\in\re^n$.  

\item[2. Cut-off to the incoming area] 
\[
\z_2(x,\x)=\i_1((\b(x,\x)-\s)/(\s-\s'))
\]
so that 
\[
\z_2(x,\x) =\begin{cases} 1, \quad &\text{if }\b(x,\x)\leq \s',\\
0,\quad &\text{if }\b(x,\x)\geq \s.
\end{cases}
\]

\item[3. Cut-off in the momentum variable]
\[
\z_3(x,\x)= \i_2\bigpare{\tfrac{|\x|^2-1}{\l(x,\x)}}, 
\quad \l(x,\x)= 2\d-\d\jap{\t(x,\x)}^{-\n}, 
\]
with $0<\n<\m$. 
\end{description}

We note $\t(x,\x)$ is decreasing along the free trajectory $x+t\hat v(\x)$
inside the cone $\Gamma_-(\d,\s,R)$, and hence $\l(x,\x)$ is also decreasing there. 
Thus $\z_3(x,\x)$ has support size decreasing along the trajectory, and $\l(x,\x)=\d$ 
at the boundary: $(x,\x)$ : $\b(x,\x)=\s$. Note also $\l(x,\x)\to 2\d$ as $t\to-\infty$ along 
the free trajectory. 

The cut-off function $\z_-(x,\x)$ satisfies the following properties: 
We denote
\[
\Gamma_-(\d,\s,R)
=\bigset{(x,\x)}{1-\d\leq |\x|^2\leq 1+\d,\b(x,\x)\leq \s, |x|\geq R}
\]
where $\d>0$, $\s\in [-1,1]$ and $R>0$.

\begin{lem}\label{lem-incoming-cut-off}
For $0<\d\ll 1$ and $0<\s'<\s<1$, there are $R_0$ and $C_0>1$ 
such that for $R\geq R_0$ there exists 
$\z_-(\cdot,\cdot)=\z_-(\d,\s,\s',R;\cdot,\cdot)\in S^{0,0}$ such that 
\[
\supp[\z_-]\subset \Gamma_-(4\d, \s,  R), 
\quad 
\z_-(x,\x)=1 \text{ on } \Gamma_-(\d,\s',C_0R), 
\]
and 
\[
\{p_2,\z_-\}(x,\x)\leq 0
\]
for all $(x,\x)\in\re^{2n}$. 
\end{lem}

\begin{figure}[t]
\centering
\includegraphics{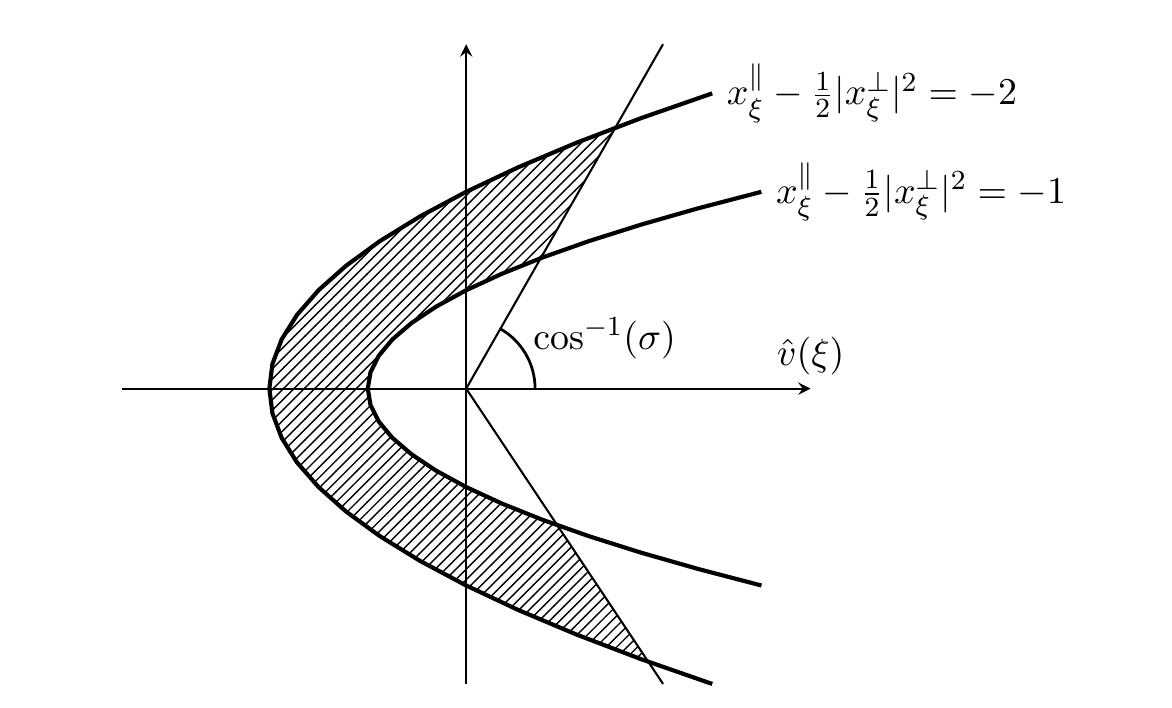}
\caption{Shaded domain : $x$-space projection of $\supp \nabla\z_1 \cap \supp \z_2$}
  \label{fig:zeta}
\end{figure}

\begin{rem}
The product $\z_-$ belongs to a good symbol class $S^{0,0}$, although $\z_1\notin S^{0,0}$. This is mainly because the projection of $\supp\nabla \z_1\cap \supp \z_2$ into the $x$-space is compact. See Figure \ref{fig:zeta}.
\end{rem}

The proof is elementary, and we give it in Appendix~\ref{app-2}. 

We need to compute an upper bound of $\{p_2,b_-\}$ 
by a nonpositive function. 
By straightforward computations, we can show 

\begin{lem}\label{lem-basic-incoming-commutator-estimate}

\noindent$(i)$ $\{p_2,\t\}+|v(\x)|\in S^{1,-\m}$ away from $\{x=0\}\cup \{\x=0\}$.

\noindent$(ii)$
Under the above assumptions, for sufficiently large $R>0$, there is $c_1>0$ such that 
\[
\{p_2,b_-\} \leq - c_1 \jap{x}^{-1} b_-. 
\]
\end{lem}

\begin{rem}
We note $b_-\in S^{0,\c}$. 
In fact $\z_-(x,\x)$ is compactly supported in $\x$, and hence 
the decay with respect to $\x$ is irrelevant.  
Since $p_2\in S^{2,0}$, we have 
$\{p_2,b_-\}\in S^{1,-1+\c}$.
\end{rem}

\begin{proof}

\noindent$(i)$ We recall 
\[
\t(x,\x)=c_0\sqrt{|x|^2-(x\cdot \hat v(\x))^2}-x\cdot\hat v(\x)
=c_0 |x_\x^\perp| -x\cdot\hat v(\x), 
\]
and hence 
\[
\pa_x \t(x,\x)= c_0\, \widehat{x_\x^\perp} -\hat v(\x),
\]
and in particular, 
\begin{equation}\label{eq-tau-gradient}
-\hat v(\x)\cdot \pa_x \t(x,\x)=1.
\end{equation}
Hence, recalling $\pa_\x p_2-v(\x)\in S^{1,-\m}$, we have 
\[
\{p_2,\t\} = \pa_x\t\cdot\pa_\x p_2-\pa_\x\t\cdot\pa_x p_2
=-|v(\x)|+S^{1,-\m}
\]
away from $\{x=0\}\cup \{\x=0\}$.

\noindent$(ii)$ We look at 
\begin{align*}
\{p_2,b_-\}&=\{p_2,\t(x,\x)^\c\}\z_-+\t(x,\x)^\c\{p_2,\z_-\}\\
&=\c\t(x,\x)^{\c-1}\{p_2,\t(x,\x)\}\z_-+\t(x,\x)^\c\{p_2,\z_-\}\\
&\leq \c\t(x,\x)^{\c-1}\{p_2,\t(x,\x)\}\z_-
\end{align*}
Thus we learn 
\begin{align*}
\{p_2,b_-\}&\leq -\c\t(x,\x)^{\c-1}(|v(\x)|+O(\jap{x}^{-\m}\jap{\x}))\z_-\\
&\leq -c_1\jap{x}^{-1} |v(\x)|\t(x,\x)^\c\z_-
\leq -c_1\jap{x}^{-1}b_-
\end{align*}
since $c|x|\leq \t(x,\x)\leq C|x|$ with some constants 
$0<c<C<\infty$, and $1-4\d\leq |\x|\leq 1+4\d$ on $\supp[\z_-]$. 
\end{proof}

\subsection{Incoming regularities} 

We denote $P_2 =\Op(p_2)=h^{-2}\Op_h(p_2)$. The sharp G{\aa}riding inequality and 
Lemma~\ref{lem-basic-incoming-commutator-estimate} imply, if we set $B_0= \Op_h(b_-)$, then 
\[
i[B_0^2, P_2] \geq  \frac{2c_0}{h} B_0 \jap{x}^{-1} B_0 + \tilde E_0, \quad \tilde E_0=\Op_h(\tilde r_0),
\]
where $\tilde r_0\in S^{-N,-2+2\c}$ with any $N\geq 0$, and $\tilde r_0$ is 
supported in $\supp[b_0]$ modulo $h^\infty S^{-\infty,-\infty}$ terms. 

Now we consider the total Hamilton operator $P$, and we denote $P=P_2+Q$ and $Q=\Op_h(q)$. 
By simple computations, we learn 
\[
q(h;x,\x)= h^{-1} \sum_{j=1}^n u_j(x)\x_j +u_0(x) +\frac14\sum_{j,k=1}^n \pa_{x_j}\pa_{x_k}g^{jk}(x), 
\]
and in particular $q\in h^{-1}S^{1,-\m}$. 
This implies the symbol of $-i[B_0^2,Q]$ is in 
$S^{0,-1-\m+2\c}$. 

Based on this observation, and by a standard semiclassical analysis argument, we can show 
the following: 

We set 
\[
0<\d_0<\d_1<\cdots <\d_\infty\ll1, \quad 0<\s_0<\s_1<\cdots<\s_\infty<1,
\]
\[
0<\s'_0<\s'_1<\cdots<\s'_\infty<1, \quad \s'_j<\s_j \quad\text{for $j=0,1,\dots$}, 
\]
and we choose 
\[
R_0>R_1>\cdots >R_\infty>0
\]
so that 
\[
\z_j(x,\x)=\z_-(\d_j,\s_j,\s'_j,R_j;x,\x)
\]
satisfies the conditions of Lemma~\ref{lem-incoming-cut-off} for all $j$. 
We may suppose the constant $c_0$ in Lemma~\ref{lem-incoming-cut-off} is 
independent of $j$, i.e., $\{p_2,b_j\}\leq -c_0\jap{x}^{-1}b_j$ for all $j$. 
We then set 
\[
b_j(x,\x)=\t(x,\x)^\c\z_j(x,\x)
\]
and 
\[
B_j =\Op_h(b_j)\in \bigcap_{k\in\re}\Op_h (S^{k,\c}).
\]

\begin{lem}\label{lem-main-ineq-negative}
For each $j=0,1,2\dots$, there is a positive constant $\a_j>0$ such that
\[
i[B_j^2, P] \geq  \frac{c_0}{h} B_j \jap{x}^{-1} B_j -\a_j B_{j+1}\jap{x}^{-1} B_{j+1}- E_j^* E_j
\]
where $\norm{E_j}=O(h^\infty)$ as $h\to 0$. 
\end{lem}

\begin{proof}
We note that the principal symbol of $ih[B_j^2,P_2]-2c_j B_j\jap{x}^{-1}B_j$ is 
\[
-\{p_2,b_j^2\}-2c_j\jap{x}^{-1}b_j^2 = 2b_j(-\{p_2,b_j\}-c_j\jap{x}^{-1}b_j), 
\]
and by Lemma~\ref{lem-basic-incoming-commutator-estimate}, 
we learn it is non-negative with suitable choice of $c_j>0$. 
We also note the symbol is in $S^{0,-1+2\c}$. 
Then, by the sharp G{\aa}rding inequality, there is 
$a_0\in S^{0,-2+2\c}$
such that 
\[
ih[B_j^2,P_2]-2c_j B_j\jap{x}^{-1}B_j \geq -h\Op_h(a_0), 
\]
and we may assume $\supp[a_0]\subset\supp[b_j]$ modulo $O(h^\infty)$ smoothing terms. 
On the other hand, the symbol $a_1$ of $-i[B_j^2,Q]$ satisfies $a_1\in S^{0,-1-\m+2\c}$, 
and also $\supp[a_1]\subset\supp[b_j]$ modulo $O(h^\infty)$ smoothing terms. 
By the construction of $b_{j+1}$, we have $(a_0+a_1) b_{j+1}^{-2}\in S^{0,-1}$. 
Using these observation and the construction of the parametrix, we can construct 
$\tilde a\in S^{0,0}$ such that
\[
\Op_h(a_0+a_1)=(\jap{x}^{-1/2}B_{j+1})^* \Op_h(\tilde a)(\jap{x}^{-1/2}B_{j+1})
\]
modulo $O(h^\infty)$ smoothing terms. Hence we have
\begin{align*}
\jap{\f,\Op_h(a_0+a_1)\f}&=\jap{(\jap{x}^{-1/2}B_{j+1})\f,\Op_h(\tilde a)(\jap{x}^{-1/2}B_{j+1})\f}+O(h^\infty)\norm{\f}^2\\
&\leq \a_j\norm{\jap{x}^{-1/2}B_{j+1}\f}^2+O(h^\infty)\norm{\f}^2
\end{align*}
where $\a_j>0$. Combining these, we conclude
\[
i[B_j^2,P] \geq \frac{2c_j}{h} B_j\jap{x}^{-1}B_j -\a_j B_{j+1}\jap{x}^{-1} B_{j+1} +O(h^\infty),
\]
which completes the proof. 
\end{proof}

Then, by \eqref{eq-basic-commutator-inequality} with $\g$, $(P-z)\g=0$, we learn 
\[
\frac{c}{2h}\bignorm{\jap{x}^{-1/2}B_j\g}^2 +2(\Im z)\norm{B_j\g}^2 
\leq \a_j\norm{\jap{x}^{-1/2}B_{j+1}\g}^2+\norm{E_j\g}^2 
\]
for each $j$, and in particular, for any $N\in \na$, 
\beq\label{eq-incoming-smoothness-iteration-step}
\norm{\jap{x}^{-1/2}B_j\g}^2 
\leq h(2\a_j/c)\norm{\jap{x}^{-1/2}B_{j+1}\g}^2+M_j h^{2N}\norm{\g}^2,  
\eeq
with some $C,M_j>0$. 
At first, setting $j=2N$ in \eqref{eq-incoming-smoothness-iteration-step}, 
we learn $\norm{\jap{x}^{-1/2}B_{2N}\g}=O(\sqrt{h})$ since $\jap{x}^{-1/2}B_{2N+1}$ 
is bounded in $L^2(\re^n)$ and $\g\in L^2(\re^n)$. Then we use this and 
\eqref{eq-incoming-smoothness-iteration-step} with $j=2N-1$, we learn 
$\norm{\jap{x}^{-1/2}B_{2N-1}\g} = O(h)$. 
Iterating this procedure $2N$ times, and we arrive at 
\[
\norm{\jap{x}^{-1/2}B_1\g}=O(h^N)
\]
for arbitrary $N\in\na$. Since $\Im z>0$, this then implies 
\beq\label{eq-incoming-smoothness}
\norm{B_0\g}=O(h^{N})
\eeq
with any $N$. 

We set $\i_3(\x)=\i_1(1-|\x|)$. Let $\s_0$, $\s_0'$ and $R_0$ as in Lemma~\ref{lem-main-ineq-negative}, 
and we denote
\[
\z^0_-(x,\x)=\z^0_-(\s_0,\s_0',R_0;x,\x) =\z_1(x/R_0,\x)\z_2(\s_0,\s_0';x,\x)\i_3(\x), 
\]
i.e., the cut-off function $\z_-(x,\x)$ without the cut-off in the momentum variable, but with a cut-off 
$\i_3(\x)$ to eliminate singularities at $\x=0$. 
We also denote 
\[
b^0(x,\x)=\t(x,\x)^\c\z^0_-(x,\x).
\]
$\z_-^0$ and $b^0$ are homogeneous of order 0 in $\x$, except for $\i_3(\x)$. We set
\[
\tilde\Gamma_-(\s,R)=\bigset{(x,\x)}{\b(x,\x)\leq \s, |x|\geq R}\subset T^*\re^n, 
\]
Then we have the following lemma.

\begin{lem}\label{lem-incoming-regularity}
Suppose $\g\in L^2(\re^n)$ and $(P-z)\g=0$ with $\Im z>0$. Then 
\begin{enumerate}
\item $\Op(b^0) \g\in H^\infty(\re^n)=\bigcap_{j=0}^\infty H^j(\re^n)$. 
\item $\jap{x}^{-\c}\Op(\z_-^0)\g\in H^\infty(\re^n)$.
\item $\mathrm{WF}(\g)\cap \tilde \Gamma_-(\s_0',R_0)=\emptyset$, 
where $\mathrm{WF}(\cdot)$ denotes the wave front set. 
\end{enumerate}
\end{lem}

\begin{proof}
The statement i) follows from \eqref{eq-incoming-smoothness} and 
the standard semiclassical characterization of the smoothness, 
or equivalently, the Besov space argument. The statement ii) follows from i), 
since $\t(x,\x)^\c\leq \jap{x}^\c$. The last statement iii) follows 
as well as i) from \eqref{eq-incoming-smoothness}  and 
the semiclassical characterization of the wave front set since $\z_-^0(x,\x)$ does not 
vanish on $\tilde \Gamma_-(\s_0',R_0)$. 
\end{proof}


\section{Overall smoothness: Propagation of singularities theorem}\label{sec-global-regularity}

Now we use the nontrapping assumption and the celebrated propagation of singularities 
theorem of H\"ormander to show that $\g$ is smooth everywhere. 

By the nontrapping condition, for any $(x_0,\x_0)\in p_2^{-1}(\{0\})$ with $\x_0\neq 0$, 
there exists $\x_-\neq 0$ such that $\eta(t;x_0,\x_0)\to \x_-$ as $t\to-\infty$, and also $y(t;x_0,\x_0)/t\to v(\x_-)$
(see \cite[$(B.1)$]{NT}), and hence we have
\[
\hat y(t;x_0,\x_0)\cdot \hat v(\eta(t;x_0,\x_0))\to -1\quad \text{as }t\to-\infty.
\]
In particular, the trajectory enters $\tilde\Gamma_-(\s_0',R_0)$ for $t\ll 0$. 
Then by the result of the last step and the propagation of singularities theorem, 
we learn $(x_0,\x_0)$ is not in the wave front set of $\g$. Thus we have the following lemma 
from Lemma~\ref{lem-incoming-regularity} and the propagation of singularities 
theorem (\cite{Ho} Theorem~23.2.9): 

\begin{lem}\label{lem-global-regularity}
Suppose $\g\in L^2(\re^n)$ and $(P-z)\g=0$ with $\Im z>0$. Then $\g\in C^\infty(\re^n)$. 
\end{lem}


\section{Outgoing estimates}\label{sec-outgoing-estimates}

\subsection{Outgoing observable}
In this section, we use the symbols $\s,\s',\s_\infty$, etc., as in Section~3, 
but here we asign different values. Let $\s_\infty\in (-1,0)$, close to $-1$, 
e.g., $\s_\infty=-9/10$. Then we set
\beq\label{eq-def-tau-outgoing}
\t(x,\x) = |x|\bigpare{\b(x,\x)-c_0\sqrt{1-\b(x,\x)^2}}, \quad c_0=\frac{\s_\infty}{\sqrt{1-\s_\infty^2}}
\eeq
for $(x,\x)$ such that $\b(x,\x)\geq \s_\infty$.

Let $\s_\infty<\s<\s'<0$, $0<\d\ll1$ and $R>0$. 
For the outgoing cut-off we can use the following construction: 
\[
\z_+(\d,\s,\s',R;x,\x)=\tilde\z_1(x/R,\x)\tilde \z_2(x,\x) \tilde \z_3(x,\x),
\]
where
\begin{align*}
&\tilde\z_1(x,\x)= \i_1\bigpare{-x_\x^\parallel-\tfrac12 |x^\perp_\x|^2+1},\\
&\tilde\z_2(x,\x)=\i_1((\s-\b(x,\x))/(\s'-\s)),\\
&\tilde\z_3(x,\x)= \i_2\bigpare{\tfrac{|\x|^2-1}{\l_+(x,\x)}}, 
\quad \l_+(x,\x)= \d_0+\d_0\jap{\t(x,\x)}^{-\n},
\end{align*}
We note that here $\t(x,\x)$ is defined by \eqref{eq-def-tau-outgoing}
and increasing along the free classical trajectory $x+tv(\x)$. 
Thus $\l_+(x,\x)$ is a decreasing function on the support of $\z_+$. 

Analogously to the incoming case, we set
\[
\Gamma_+(\d,\s,R)
=\bigset{(x,\x)}{1-\d\leq |\x|^2\leq 1+\d,\b(x,\x)\geq \s, |x|\geq R}
\]
where $\d>0$, $\s\in [-1,1]$ and $R>0$. Then we have

\begin{lem}\label{lem-outgoing-cut-off}
For $0<\d\ll 1$ and $-\s_\infty<\s<\s'<0$, there is $R_0$ and $C_0>0$ 
such that for $R\geq R_0$ there exists 
$\z_+(\cdot,\cdot)=\z_+(\d,\s,\s',R;\cdot,\cdot)\in S^{0,0}$ such that 
\[
\supp[\z_+]\subset \Gamma_+(4\d, \s,  R), 
\quad 
\z_+(x,\x)=1 \text{ on } \Gamma_+(\d,\s',C_0R), 
\]
and 
\beq\label{eq-outgoing-inequality}
\{p_2,\z_+\}(x,\x)\leq \rho(x,\x), \quad (x,\x)\in\re^{2n},
\eeq
where $\rho=\rho(\d,\s,\s',R:\cdot,\cdot)\in S^{0,-1+\c}$ such that
\[
\supp[\rho]\subset \bigset{(x,\x)}{1-4\d\leq |\x|^2\leq 1+4\d, |x|\leq C_0R\text{ or }
\s \leq \b(x,\x)\leq \s'}. 
\]
\end{lem}

The proof is essentially the same as Lemma~\ref{lem-incoming-cut-off}, and we 
sketch the proof in Appendix~\ref{app-2}.

We set 
\[
0<\d_0<\d_1<\cdots <\d_\infty\ll 1, \quad 0>\s_0>\s_1>\cdots>\s_\infty
\]
\[
0>\s'_0>\s'_1>\cdots>\s'_\infty>\s_\infty, \quad \s_j<\s'_j \quad\text{for $j=0,1,\dots$}, 
\]
and we choose 
\[
R_0>R_1>\cdots >R_\infty>0
\]
so that 
\[
\z_j^+(x,\x)=\z_+(\d_j,\s_j,\s'_j,R_j;x,\x)
\]
satisfies the conditions of Lemma~\ref{lem-outgoing-cut-off} for all $j$. 
Then we set 
\[
b_j^+(x,\x) = \t(x,\x)^\c\z_j^+(x,\x)
\quad 
\text{and}
\quad
B_j^+=\Op_h(b_j^+). 
\]
We denote $\rho(x,\x)$ in Lemma~\ref{lem-outgoing-cut-off} with the constants $\d_j,\s_j,\s_j',R_j$
by $\rho_j(x,\x)$. 
Then, as well as Lemma~\ref{lem-main-ineq-negative}, we similarly have
\[
i[(B_j^+)^2, P] \geq  \frac{c_0}{h} B_j^+ \jap{x}^{-1} B_j^+ -\a_j  B_{j+1}^+ \jap{x}^{-1} B_{j+1}^+ -S_j - E_j^*E_j
\]
where $S_j$ is an $h$-pseudodifferential operator in $\Op(h^{-1}S^{0,-1+\c})$ 
with the principal symbol $h^{-1}\rho_j=h^{-1}\rho(\d_j,\s_k,\s'_j,R_j;\cdot,\cdot)$ and it has the same support 
as $\rho_j$, and $\norm{E_j}=O(h^\infty)$. 
Thus, we again learn, for any $N\in \na$, 
\[
\frac{c}{2h}\bignorm{\jap{x}^{-1/2}B_j^+\g}^2 +2(\Im z)\norm{B_j^+\g}^2 
\leq \a_j\norm{\jap{x}^{-\c}B^+_{j+1}\g}^2+\jap{\g,S_j\g}+\norm{E_j\g}^2 
\]
for each $j$. We note that the support of $\rho_j$ is contained in $\Gamma_-(4\d_j,\s_j,\s_j',R)$ 
outside a compact set in $x$-space, and hence by Lemmas~\ref{lem-incoming-regularity} and 
\ref{lem-global-regularity}, we learn $\jap{\g,S_j\g}=O(h^N)$ with any $N$ as $h\to 0$. 
In particular, we have 
\beq\label{eq-outgoing-smoothness-iteration-step}
\norm{\jap{x}^{-1/2}B_j^+\g}^2 
\leq Ch \a_j\norm{\jap{x}^{-\c}B_{j+1}^+\g}^2+C_j h^{2N}\norm{\g}^2. 
\eeq
Now using the same iteration step as in the incoming case, we have:  
\beq\label{eq-outgoing-smoothness}
\norm{B_0^+\g}=O(h^N)
\eeq
for arbitrary $N\in\na$. Now we have the regularity theorem in the outgoing case, as well 
as the incoming case. 

Let $\s_0$, $\s_0'$ and $R_0$ as above. Analogously to the incoming case, we denote
\begin{align*}
&\z^0_+(x,\x)=\z^0_+(\s_0,\s_0',R_0;x,\x) =\tilde\z_1(x/R_0,\x)\tilde\z_2(\s_0,\s_0',;x,\x)\i_3(\x), \\
&b^0_+(x,\x)=\t(x,\x)^\c\z^0_+(x,\x), \\
&\tilde\Gamma_+(\s,R_0)=\bigset{(x,\x)}{\b(x,\x)\geq \s, |x|\geq R}\subset T^*\re^n.
\end{align*}

\begin{lem}\label{lem-outgoing-regularity}
Suppose $\g\in L^2(\re^n)$ and $(P-z)\g=0$ with $\Im z>0$. Then 
\begin{enumerate}
\item $\Op(b^0_+) \g\in H^\infty(\re^n)$. 
\item $\jap{x}^{-\c}\Op(\z_+^0)\g\in H^\infty(\re^n)$.
\end{enumerate}
\end{lem}


\section{Proof of Theorem~1}\label{sec-proof-of-main-theorem}

Combining results of Lemmas~\ref{lem-incoming-regularity}, \ref{lem-global-regularity}
and \ref{lem-outgoing-regularity}, we learn $\g\in H^{-N,\c}(\re^n)$ with any $N\in\na$
and sufficiently small  $\c>0$, 
provided $(P-z)\g=0$ and $\g\in L^2(\re^n)$, $\Im\, z>0$. Then, by Lemma~\ref{lem-the-first-reduction}, 
we conclude $\g=0$
and hence $\Ker(P^*-z)=\{0\}$. Similarly we can prove $\Ker(P^*-z)=\{0\}$ when 
$\Im\, z<0$. We note we have used the \emph{incoming} nontrapping condition in the above 
argument, but we use the \emph{outgoing} nontrapping condition when $\Im\, z<0$. \qed 

\begin{appendices}

\section{Proof of Lemma~\ref{lem-the-first-reduction}}
\label{app-0}


Suppose $\g$ satisfies the conditions of Lemma~\ref{lem-the-first-reduction}. 
At first we note that if $\f\in H^1(\re^n)$ and $P\f\in L^2(\re^n)$, then by the definition of the 
distributional derivative, we learn 
\begin{multline*}
\jap{\f, P\f} = \sum_{j,k=1}^n \int g^{jk}(x)\overline{D_j\f(x)}D_k\f(x)dx \\
+\Re\biggpare{\sum_{j=1}^n \int u_j \overline{\f(x)}D_j\f(x)dx} +\int u_0|\f(x)|^2 dx \in\re.
\end{multline*}
We choose a smooth function $\chi\in C_0^{\infty}(\re^n;[0,1])$ such that $\chi(x,\x)=1$ for 
$|x|\leq 1$. We set $X_R\f(x)=\chi(x/R)\f(x)$ for $R>0$ and $\f\in L^2(\re^n)$. 
Then $X_R\g\in H^1(\re^n)$ for each $R>0$, and hence we learn 
\[
\Im \jap{X_R\g,(P-z)X_R\g} = -\Im\, z\norm{X_R\g}^2. 
\]
On the other hand, we have
\begin{align*}
\jap{X_R\g,(P-z)X_R\g} &=\jap{X_R\g, X_R(P-z)\g}+\jap{X_R\g,[P,X_R]\g}\\
&=\jap{X_R\g,[P,X_R]\g}.
\end{align*}
It is easy to observe that $[P,X_R]$ is a first order differential operator with the coefficients 
uniformly bounded by $C\jap{x}^{-1}$, and converges to 0 pointwise as $R\to\infty$. 
Thus $[P,X_R]\g$ is bounded by an $L^2$ function, and then by the 
dominated convergence theorem, we have $\jap{X_R\g,[P,X_R]\g}\to 0$ as $R\to\infty$. 
Now we conclude 
\[
-\Im\, z\norm{\g}^2 =\lim_{R\to\infty}\bigpare{-\Im\, z\norm{X_R\g}^2} =\lim_{R\to\infty}\jap{X_R\g,[P,X_R]\g}=0,
\]
and thus $\g=0$. 
\qed

\section{Proof of the basic commutator estimate}
\label{app-1}

In this appendix, we prove a basic inequality used in Subsection~\ref{subseccommutator}.
More precisely, we show
\begin{align}\label{commass}
i[B^*B, P] \geq \frac{c}{h}B^* \jap{x}^{-1} B -\tilde B^* \jap{x}^{-1}\tilde B -E^*E,
\end{align}
implies
\begin{align}
&\frac{c}{2h}\bignorm{\jap{x}^{-1/2}B\f}^2 +2(\Im z)\norm{B\f}^2\nonumber \\
&\qquad \leq \frac{2h}{c}\bignorm{\jap{x}^{1/2}B(P-z)\f}^2 +\norm{\jap{x}^{-1/2}\tilde B\f}^2
+\norm{E\f}^2,\label{basiccom2}
\end{align}
where $\norm{\cdot}=\norm{\cdot}_{L^2(\re^n)}$. 

At first, we prove \eqref{basiccom2} for $\f\in\mathcal{S}(\re^n)$. 
If $\f\in\mathcal{S}(\re^n)$, we have 
\begin{align*}
i\jap{\f,[B^*B,P]\f} &=
i\jap{\f, (B^*B P-PB^*B)\f} \\
&= i(\jap{B\f,B(P-z)\f}-\jap{B(P-\bar{z})\f,B\f})\\
&= -2\Im(\jap{B\f,B(P-z)\f}) -2(\Im z)\norm{B\f}^2\\
&\leq 2\norm{\jap{x}^{-1/2} B\f}\,\norm{\jap{x}^{1/2} B(P-z)\f} 
-2(\Im z)\norm{B\f}^2.
\end{align*}
On the other hand, we have
\begin{align*}
&\frac{c}{h}\jap{\f,B\jap{x}^{-1}B\f} -\jap{\f,\tilde B^*\jap{x}^{-1}\tilde B\f}-\jap{\f,E^*E\f}\\
&\qquad = \frac{c}{h}\norm{\jap{x}^{-1/2}B\f}^2-\norm{\jap{x}^{-1/2}\tilde B\f}^2-\norm{E\f}^2.
\end{align*}
Combining them with our assumption (\ref{commass}), we learn 
\begin{align*}
&\frac{c}{h}\norm{\jap{x}^{-1/2}B\f}^2+2(\Im z)\norm{B\f}^2\\
&\qquad \leq 2\norm{\jap{x}^{-1/2} B\f}\,\norm{\jap{x}^{1/2} B(P-z)\f} +\norm{\jap{x}^{-1/2}\tilde B\f}^2 +\norm{E\f}^2.
\end{align*}
Now we use the elementary bound:
\[
\norm{\jap{x}^{-1/2} B\f}\,\norm{\jap{x}^{1/2} B(P-z)\f}
\leq \frac{c}{4h}\norm{\jap{x}^{-1/2} B\f}^2+ \frac{h}{c}\norm{\jap{x}^{1/2} B(P-z)\f}^2
\]
in the right hand side, and we obtain \eqref{basiccom2} for $\f\in \mathcal{S}(\re^n)$.

In applications, we use \eqref{basiccom2} for $\f\in L^2(\re^n)$ 
such that $(P-z)\f\in H^{0,1/2+\c}$, and we need to 
show the inequality extends to such functions. Since $B, \tilde B, E\in \bigcap_{m\in \re}\Op S^{m,\c}$, it is easy to observe that \eqref{basiccom2} is extended to 
$\f\in \bigcap_{\ell\in \re} H^{0,\ell}$.  

Now let $A$ be one of the operators $\jap{x}^{-1/2}B$, $B$, $\jap{x}^{1/2}B(P-z)$, $\jap{x}^{-1/2}\tilde B$
and $E$. Let $X_R$ be the operator used in the last Appendix. Then $[A,X_R]$ is a pseudodifferential 
operator with the symbol which is bounded in $S^{0,-1/2+\c}$ and supported in 
$\supp[\nabla X_R]\subset\{|x|\geq R\}$. 
These imply 
\[
\norm{[A,X_R]}_{L^2\to L^2}\leq CR^{-1/2+\c}\to 0, \quad \text{as }R\to\infty
\]
by the $L^2$-boundedness theorem for pseudodifferential operators. 
Using this, and since $X_R\f\in \bigcap_{\ell\in \re} H^{0,\ell}(\re^n)$ if $\f\in L^2(\re^n)$,  we have 
\begin{align*}
&\frac{c}{2h}\bignorm{\jap{x}^{-1/2}B\f}^2 +2(\Im z)\norm{B\f}^2 \\
&\qquad =\lim_{R\to\infty}\biggpare{\frac{c}{2h}\bignorm{X_R\jap{x}^{-1/2}B\f}^2 +2(\Im z)
\norm{X_R B\f}^2} \\
&\qquad \leq \lim_{R\to\infty}\biggpare{\frac{2h}{c}\bignorm{X_R\jap{x}^{1/2}B(P-z)\f}^2 
+\norm{X_R\jap{x}^{-1/2}\tilde B\f}^2
+\norm{X_RE\f}^2}\\
&\qquad = \frac{2h}{c}\bignorm{\jap{x}^{1/2}B(P-z)\f}^2 +\norm{\jap{x}^{-1/2}\tilde B\f}^2
+\norm{E\f}^2,
\end{align*}
provided $\f\in L^2(\re^n)$ and $(P-z)\f\in H^{0,1/2+\c}$. \qed

\section{Proof of Lemmas~\ref{lem-incoming-cut-off} and \ref{lem-outgoing-cut-off}}
\label{app-2}

\begin{proof}[Proof of Lemma~\ref{lem-incoming-cut-off}]
It suffices to prove that for each $j=2,3$,
\begin{align*}
\{p_2,\z_1(x/R,\x)\} \leq 0,\quad \{p_2,\z_j(x,\x)\}\leq 0
\end{align*}
on $\supp [\z_1(\cdot/R,\cdot)\z_2\z_3]$ for sufficiently large $R$.

Throughout this proof, we denote $\d=\s-\s'>0$ for simplicity. 
At first, we consider the estimate for $\z_1(x/R,\x)$.
 We note 
\[
\supp[\z_1]\subset \bigset{(x,\x)}{x_\x^\parallel\leq -1+\tfrac12|x_\x^\perp|^2}
\subset\bigset{(x,\x)}{|x|\geq 1},
\]
and 
\[
\supp[\pa_{(x,\x)}\z_1]\subset \bigset{(x,\x)}{-2+\tfrac12|x_\x^\perp|^2
\leq x_\x^\parallel\leq -1+\tfrac12|x_\x^\perp|^2}.
\]
Moreover, 
\[
\hat v(\x)\cdot\pa_x \z_1(x,\x) =\i_1'\bigpare{x_\x^\parallel-\tfrac12 |x^\perp_\x|^2+1}\leq 0.
\]
Since $\pa_\x p_2=v(\x)+O(|\x|\jap{x}^{-\m})$, we have 
\[
\pa_\x p_2\cdot\pa_x \z_1(x/R,\x) 
= R^{-1} \i_1'\bigpare{(x/R)_\x^\parallel-\tfrac12 |(x/R)^\perp_\x|^2+1}\cdot\bigpare{1+O(R^{-\m})}|\x|. 
\]
Similarly, since $\pa_x p_2=O(|\x|^2\jap{x}^{-1-\m})$ and $\z_1(x,\x)$ is homogeneous in $\x$, we learn 
\[
\pa_x p_2\cdot \pa_\x\z_1(x/R,\x)
= \i_1'\bigpare{(x/R)_\x^\parallel-\tfrac12 |(x/R)^\perp_\x|^2+1}\cdot O(R^{-1-\m})|\x|.
\]
These imply 
\begin{align*}
\{p_2,\z_1(x/R,\x)\} 
= R^{-1}\i_1'\bigpare{(x/R)_\x^\parallel-\tfrac12 |(x/R)^\perp_\x|^2+1}\cdot \bigpare{1+O(R^{-\m})}|\x|
\leq 0
\end{align*}
on $\supp [\z_1(x/R,\x)\z_2(x,\x)\z_3(x,\x)]$ for sufficiently large $R$.

Next, we deal with the estimate for $\z_2$.
We recall $\z_2$ is homogenous in $(x,\x)$, and we note 
\begin{align*}
\pa_x\z_2(x,\x) &= (\pa_x\b(x,\x)) \i_1'((\b(x,\x)-\s)/\d))/\d\\
&=|x|^{-1}(\hat v(\x)-\hat x \b(x,\x))\cdot \i_1'((\b(x,\x)-\s)/\d)/\d, 
\end{align*}
and in particular
\begin{align*}
\hat v(\x)\cdot\pa_x\z_2(x,\x) &=|x|^{-1}(1-\b(x,\x)^2) \i_1'((\b(x,\x)-\s)/\d)/\d \\
&\leq \d^{-1}|x|^{-1} (1-\s^2)\i_1'((\b(x,\x)-\s)/\d) \leq 0. 
\end{align*}
Similarly to the argument for $\z_1$, if $|x|\geq R$ then we have 
\[
\pa_\x p_2\cdot \pa_x \z_2 =\d^{-1}|x|^{-1} (1-\s^2)\i_1'((\b(x,\x)-\s)/\d) \cdot 
(1+O(R^{-\m}))|\x|
\]
and 
\[
\pa_x p_2\cdot\pa_\x \z_2(x,\x)= \i_1'((\b(x,\x)-\s)/\d)\cdot O(|x|^{-1-\m}|\x|).
\]
Combining these, we have 
\begin{align*} 
\{p_2,\z_2\} =\d^{-1}|x|^{-1} (1-\s^2)\i_1'((\b(x,\x)-\s)/\d) \cdot 
(1+O(R^{-\m}))|\x|\leq 0
\end{align*}
on $\supp [\z_1(x/R,\x)\z_2(x,\x)\z_3(x,\x)]$.

Finally, we consider the estimate for $\z_3$.
We now note $\t(x,\x)$ is the length of the line segment 
$\bigset{x+t\hat v(\x)}{t\geq 0}$ 
inside $\bigset{(x,\x)}{\b(x,\x)\leq \s_\infty}$.  
We recall 
\[
\t(x,\x)=c_0\sqrt{|x|^2-(x\cdot \hat v(\x))^2}-x\cdot\hat v(\x)
=c_0 |x_\x^\perp| -x\cdot\hat v(\x), 
\]
and hence 
\[
\pa_x \t(x,\x)= c_0\, \widehat{x_\x^\perp} -\hat v(\x),
\]
and in particular, 
\begin{equation}\label{eq-App-B-3}
-v(\x)\cdot \pa_x \t(x,\x)=|v(\x)|.
\end{equation}
We also note 
\[
c_1|x|\leq \t(x,\x)\leq C_1|x| \quad\text{for }(x,\x)\in\supp[\z_2]
\]
with some $0<c_1<C_1$. 
We also note
\[
\pa_\x\z_3(x,\x) = \i_2'\bigpare{\tfrac{|\x|^2-1}{\l(x,\x)}} \biggpare{\frac{2\x}{\l(x,\x)} -(|\x|^2-1)\frac{\pa_\x\l(x,\x)}{\l(x,\x)^2}}
\]
where $(\cdots)$ is smooth and uniformly bounded on the support of $\i_2'(\cdots)$. 
On the other hand, 
\begin{align*}
\pa_x \z_3(x,\x)&= -\i_2'\bigpare{\tfrac{|\x|^2-1}{\l(x,\x)}}(|\x|^2-1) \frac{\pa_x\l(x,\x)}{\l(x,\x)^2}\\
&=-\i_2'\bigpare{\tfrac{|\x|^2-1}{\l(x,\x)}}\frac{|\x|^2-1}{\l(x,\x)^2}\frac{\n\d_0\t(x,\x)\pa_x\t(x,\x)}{\jap{\t(x,\x)}^{2+\n}},
\end{align*}
and in particular, by \eqref{eq-App-B-3}, we have 
\[
v(\x)\cdot\pa_x \z_3(x,\x) = \i_2'\bigpare{\tfrac{|\x|^2-1}{\l(x,\x)}}\frac{|\x|^2-1}{\l(x,\x)^2}\frac{\n\d_0\t(x,\x)|v(\x)|}{\jap{\t(x,\x)}^{2+\n}}. 
\]
This also implies 
\[
\pa_\x p_2(x,\x)\cdot\pa_x\z_3(x,\x)=
\i_2'\bigpare{\tfrac{|\x|^2-1}{\l(x,\x)}}\frac{|\x|^2-1}{\l(x,\x)^2}\frac{\n\d_0\t(x,\x)|v(\x)|}{\jap{\t(x,\x)}^{2+\n}}(1+O(\jap{x}^{-\m})).
\]
Noting $\i'_2(t)t\leq 0$ for $t\in\re$ and $|t|\geq 1$ on 
$\supp[\i_2'(t)]$, we learn that 
\[
\pa_\x p_2(x,\x)\cdot\pa_x\z_3(x,\x) \leq - c_2 \bigabs{\i_2'\bigpare{\tfrac{|\x|^2-1}{\l(x,\x)}}}\jap{x}^{-1-\n}
\]
with some $c_2>0$ on $\supp[\z_2]$. 
On the other hand, using $\pa_x p_2(x,\x)=O(|\x|^2\jap{x}^{-1-\m})$ again, we have 
\[
\pa_x p_2(x,\x)\cdot \pa_\x\z_3(x,\x)= \i_2'\bigpare{\tfrac{|\x|^2-1}{\l(x,\x)}}\times O(|\x|^3\jap{x}^{-1-\m}). 
\]
Since $|\x|$ is bounded on $\supp \z_3$, these imply
\[
\{p_2,\z_3\} \leq -c_2\bigabs{\i_2'\bigpare{\tfrac{|\x|^2-1}{\l(x,\x)}}}(\jap{x}^{-1-\n}-C\jap{x}^{-1-\m})\leq 0
\]
on $\supp [\z_1(x/R,\x)\z_2(x,\x)\z_3(x,\x)]$ with sufficiently large $R$.
\end{proof}

\begin{proof}[Proof of Lemma~\ref{lem-outgoing-cut-off}]
At first, we note if we set 
\[
\rho(x,\x)=\{p_2,\tilde\z_1(x,\x)\tilde\z_2(x,\x)\},
\]
then $\rho$ satisfies the properties of the lemma, and it suffices to show $\{p_2,\tilde\z_3\}$
is nonpositive on the support to prove the inequality \eqref{eq-outgoing-inequality}. 
The computation is almost identical to the one in the proof of Lemma~\ref{lem-incoming-cut-off} 
above, but we remark necessary changes. 
Even though the definition of $\l_+(x,\x)$ is different from $\l(x,\x)$, we have the same derivative 
formula:
\[
\pa_x\l_+(x,\x) =-\n\d_0\frac{\tau(x,\x)\pa_x\tau(x,\x)}{\jap{\tau(x,\x)}^{\n+2}},
\]
and we have the same bound eventually: 
\[
\pa_\x p_2(x,\x)\cdot\pa_x\tilde\z_3(x,\x) \leq - c_2 \bigabs{\i_2'\bigpare{\tfrac{|\x|^2-1}{\l(x,\x)}}}\jap{x}^{-1-\n}.
\]
The rest of the computation is carried out without changes to conclude 
$\tilde\z_1\tilde\z_2\{p_2,\tilde\z_3\}\leq 0$ with sufficiently large $R$. 
\end{proof}

\end{appendices}

\end{document}